\documentclass[a4paper,11pt]{article}
\usepackage{jheppub} 
\usepackage{lineno}
\usepackage{mathrsfs}
\usepackage{natbib}
\usepackage[utf8]{inputenc}
\usepackage{amsmath}
\usepackage{comment}
\usepackage[compat=1.1.0]{tikz-feynman}
\tikzfeynmanset{warn luatex=false}
\usepackage{feynmp-auto}
\usepackage{tikz}
\usepackage{subcaption}
\def\ie{{\textit{i.e.}}}

\arxivnumber{1234.56789} 

\title{\boldmath Gravitational memory effects in Tachyon gravity}

\author[a]{Pouneh Safarzadeh Ilkhchi,\note{Corresponding author.}}
\author[a]{Amin Rezaei Akbarieh}
\author[b,c]{and Ali Seraj}
\affiliation[a]{Faculty of Physics, University of Tabriz, Tabriz 51666-16471, Iran,}
\affiliation[b]{School of Quamtum Physics and Matter, Institute for Research in Fundamental Sciences (IPM), P.O.Box 19395-5531, Tehran, Iran}
\affiliation[c]{Leuven Gravity Institute, KU Leuven,
Celestijnenlaan 200D box 2415, 3001 Leuven, Belgium}

\emailAdd{p.safarzadeh@tabrizu.ac.ir, am.rezaei@tabrizu.ac.ir, ali\_seraj@ipm.ir}

\abstract{
In this paper, we explore how a massless Tachyon field, when non-minimally coupled to gravity, affects gravitational memory, soft theorems, and symmetries near null infinity. We set up the model using both the Jordan and Einstein frames, work out the equations of motion and study the behavior of fields at large distances with the Bondi-Sachs gauge. The Tachyon adds a ``breathing'' polarization mode to the gravitational radiation, which modifies the Bondi mass loss formula and leaves a trace in the displacement memory effect.  We also derived the leading and subleading soft factors, involving the radiation of a soft graviton or a soft scalar.}

\begin{document}
\maketitle
\flushbottom

\section{Introduction}
One of the significant predictions of general relativity (GR) is the generation and propagation of gravitational wave (GW), with no Newtonian counterparts. GWs are created by dynamical matter, such as colliding black holes or neutron stars. Like any other wave, GWs have a frequency spectrum, whose low-frequency behavior leads to interesting universal features in the late-time behavior of the waveform such as the memory~\cite{Zeldovich:1974gvh,Braginsky:1985vlg,braginsky1987gravitational,Christodoulou:1991cr,Thorne:1992sdb} and tail effects. These have obervable consequences such as a permanent displacement of a test masse in an inertial reference frame~\cite{Zeldovich:1974gvh} and a kick effect on moving test masses~\cite{Zhang:2017geq,Zhang:2017rno,Flanagan:2018yzh}. There are also subleading effects such as a permanent change in the orientation of a free gyroscope~\cite{Seraj:2021rxd,Seraj:2022qyt} or a phase shift in the interference pattern of a Sagnac interferometer~\cite{Pasterski:2015zua,Seraj:2022qqj}.
Late-time effects are not only interesting because of their observable signatures but also due to their relation with the universal infrared (IR) properties of gravity, which also sheds light on its symmetry structure~\cite{Strominger:2017zoo,DeLuca:2024bpt}.\\
Gravitational radiation crucially depends on the theory of gravity. The study of GWs in modified theories of gravity forms an active field of research, as it allows to constrain these theories by the observations (see~\cite{Barack:2018yly} and references therein). In modified theories of gravity with extra massless degrees of freedom, one expects to have additional memory effects. This has been tested in various modified gravity theories and in particular scalar-tensor theories \cite{Hou:2021oxe,Bernard:2022noq,Gorji:2022hyy,Heisenberg:2023prj,Trestini:2023khz,BenAchour:2024zzk,Hou:2020tnd,Tahura:2020vsa,Seraj:2021qja,Tahura:2025ebb}. In this framework, in addition to the tensor memory similar to GR, a new “breathing memory” is formed, which is related to the permanent shift in the scalar field \citep{Koyama:2020vfc}.The latter causes  a uniform expansion or contraction of a ring of test particles, hence the name breathing. Unlike tensor memory, this new effect is not directly related to boundary BMS symmetries but could be related to the asymptotic gauge symmetries of a dual formulation of the scalar~\cite{Seraj:2021qja}.\\
In this paper, we investigate the asymptotic structure of the scalar Tachyon model at null infinity (see \citep{Bose:2009zzd,Calcagni:2006ge,Sen:2002nu} for various aspects of this model in other contexts). We will derive the gravitational memory as well as the soft theorems in this model and establish their consistency, correcting some previous confusion in the literature. We will also derive the flux-balance equations in this model and emphasize how the scalar enhances the fluxes.\\
The paper is structured as follows: Section \ref{sec2} introduces the Tachyon model, deriving its field equations in Jordan and Einstein frames and its asymptotic behavior in the Bondi-Sachs gauge. Section \ref{sec3} derives flux-balance equations at null infinity. Section \ref{sec4} derives the memory effects from balance equations and explains how they induce a permanent displacement on nearby geodesics. Section \ref{sec5} examines soft theorems, covering soft graviton and scalar factors with and without matter fields. Section \ref{sec6} summarizes the findings and proposes future research directions.

\section{Non-minimally coupled Tachyon}\label{sec2}
The scalar Tachyon field, which has an unusual term of kinetic energy, is a model that is commonly studied in both string theory and cosmology \cite{Sen:2002nu, Fairbairn:2002yp,Sen:1998sm,Arutyunov:2000pe,RezaeiAkbarieh:2018ijw}. In this paper, we consider a massless Tachyon field that is non-minimally coupled to gravity \cite{Nozari:2013mba,Piao:2002nh}. The action of the proposed model in the Jordan framework is written as
\begin{equation}\label{action}
S = \frac{1}{16\pi G} \int d^4 x \sqrt{-g} \left[ \varphi R - f(\varphi) (\sqrt{1 + g^{\mu\nu}\partial_\mu \varphi \partial_\nu \varphi}-1) \right],
\end{equation}
where $\varphi$ denotes the massless Tachyon field, $R$ is the Ricci scalar, $g$ is the determinant of the metric $g_{\mu\nu}$, and $f(\varphi)$ is an arbitrary function of $\varphi$. Variation with respect to $g_{\mu\nu}$ yields the field equations
\begin{equation}\label{EqEinstein}
\varphi G_{\mu\nu} = \frac{f(\varphi)}{2} \left(g_{\mu\nu} [1-\sqrt{1 + X}]  +\frac{2 \partial_\mu \varphi \partial_\nu \varphi}{{(1 + X)^{3/2}}} \right)+(\nabla_\mu\nabla_\nu- g_{\mu\nu} \square) \varphi ,
\end{equation}
where $G_{\mu\nu}$ is the Einstein tensor, $\square = g^{\alpha\beta} \nabla_\alpha \nabla_\beta$ and the kinetic term is defined as $X = g^{\alpha\beta} \partial_\alpha \varphi \partial_\beta \varphi$. By varying the action with respect to $\varphi$, we obtain
\begin{equation} \label{EqScalar}
  \nabla_\mu \left( \frac{f(\varphi) \partial^\mu \varphi}{(1+X)^{3/2}} \right)- f'(\varphi) \big(\sqrt{1 + X} - 1 \big) + R = 0.
\end{equation}
By applying a conformal transformation to the metric~\cite{Piao:2002nh}, $\tilde{g}_{\mu\nu} = \varphi g_{\mu\nu}$, the action in the Einstein frame is written as
\begin{equation}\label{intractionaction}
S = \frac{1}{16\pi G} \int d^4 x \sqrt{-\tilde{g}} \left[ \tilde{R}  +\frac{3}{2 \varphi} \tilde{g}^{\mu\nu} \partial_\mu \varphi \partial_\nu \varphi - \frac{f(\varphi)}{\varphi^2} \left( \sqrt{1 + \varphi \tilde{g}^{\mu\nu} \partial_\mu \varphi \partial_\nu \varphi} - 1 \right) \right],
\end{equation}
where $\tilde{R}=\tilde{R}[\tilde{g}]$ is the Ricci scalar in the Einstein frame. The gravitational sector is now standard, but the scalar field acquires an additional kinetic contribution from the transformation. By varying action (\ref{intractionaction}) with respect to $\tilde{g}_{\mu\nu}$, we derive the equations of motion in the Einstein frame that yield $\tilde{G}_{\mu\nu}=8\pi G \tilde{T}_{\mu\nu} $ with
\begin{equation}
\tilde{T}_{\mu\nu} = \Big( \frac{3}{4 \varphi} - \frac{f(\varphi)}{\varphi \sqrt{1 + \varphi {\tilde X}}} \Big) \partial_\mu \varphi \partial_\nu \varphi + \left( \frac{f(\varphi)}{2 \varphi^2} (\sqrt{1 + \varphi \tilde{X}}-1) - \frac{3}{2 \varphi} \tilde{X} \right) \tilde{g}_{\mu\nu},
\end{equation}
where $\tilde{X} = \tilde{g}^{\mu\nu} \partial_\mu \varphi \partial_\nu \varphi$.
Varying action (\ref{intractionaction}) with respect to $\varphi$, we obtain
\begin{equation}
	-\frac{3}{\varphi} \tilde{\square} \varphi +\frac{3}{2\varphi^2}\tilde{X}-\frac{f'(\varphi)}{\varphi^2} \left( \tilde{Y} - 1 \right) + \frac{2f(\varphi)}{ \varphi^3} \left( \tilde{Y} - 1 \right) - \frac{f(\varphi)}{2 \varphi^2} \frac{\tilde{X}}{\tilde{Y}} +\tilde{\nabla}_\mu \left( \frac{f(\varphi) \partial^\mu \varphi}{\varphi \tilde{Y}} \right) = 0,
\end{equation}
where $\tilde{Y} = \sqrt{1 + \varphi \tilde{X}}$. In the following, we will use this framework to calculate the gravitational memory effect in the Bondi-Sachs gauge in asymptotically flat spacetime, focusing on the perturbations caused by the massless Tachyon field.

\subsection{Bondi-Sachs formulation}

To explore the dynamics of gravitational radiation and the asymptotic properties of spacetime in this model, we adopted the Bondi-Sachs coordinate system, a framework well suited for studying isolated systems emitting waves to future null infinity $\mathscr{I}^{+}$ \cite{Madler:2016xju}. This coordinate system, consisting of a retarded time $u$, a radial coordinate $r$, and a coordinate system $x^A \,(A=2,3)$  on the two-sphere, facilitates the analysis of outgoing radiation and the symmetry properties of asymptotically flat spacetimes. The general form of the metric in the Bondi-Sachs gauge is expressed as
\begin{equation}\label{bondimetric}
ds^2 = - \frac{V}{r} e^{2\beta} du^2 - 2 e^{2\beta} du dr + r^2 h_{AB} (dx^A - U^A du)(dx^B - U^B du),
\end{equation}
where $\beta$, $V$, $U^A$, and the symmetric  $h_{AB}$ are metric functions dependent on all coordinates.
A key feature of the Bondi-Sachs gauge is that the hypersurfaces of constant $u$ are null, and the radial coordinate $r$ aligns with the outgoing null rays. These are ensured by the gauge conditions $g_{rr} = 0$ and $g_{rA} = 0$ already incorporated in \eqref{bondimetric}. As a fourth gauge condition, we impose the following determinant condition which ensures that the $r$ is the areal distance with respect to the metric $\tilde{g}$ in the Einstein frame,
\begin{equation}\label{determinant}
\det(h_{AB}) = \varphi^{-2}  {\gamma}\,,
\end{equation}
where ${\gamma}=\det(\gamma_{AB})$ is the determinant of a fixed metric $\gamma_{AB}(x^A)$ on the round sphere.
One could also impose the standard Bondi determinant condition $\det(h_{AB}) = {\gamma}$. The reason we prefer \eqref{determinant} here is that it collects all the propagating modes in the transverse part of the metric, as we will see explicitly below.

Assuming analyticity of the fields at $\mathscr{I}^{+}$, we perform the following expansion in  $1/r$ in the $r \to \infty$ limit
\begin{equation}\label{expansion}
   \varphi=\sum_{n=0}^\infty{\frac{\varphi_n}{r^n}}, \;\;\; \beta=\sum_{n=0}^\infty{\frac{\beta_n}{r^n}}, \;\;\;V=-r+\sum_{n=0}^\infty{\frac{V_n}{r^n}},\;\;\;U^A=\sum_{n=2}^\infty{\frac{U^A_n}{r^n}},\;\;\;h_{AB}=\gamma_{AB}+\sum_{n=1}^\infty{\frac{C_{AB}^{(n)}}{r^{n}}},
   \end{equation}
and we Taylor expand the function $f(\varphi)$ around $\varphi_0$,
$$f(\varphi)=f_0+\frac{1}{r}f^{'}_0\varphi_1+\frac{1}{r^2}(f^{'}_0\varphi_2+\frac{1}{2}f^{''}_0\varphi_1\big)+\cdots\,.$$
All coefficients in these equations are a priori functions of $(u, x^A)$, except $\gamma_{AB}$ which is assumed to be time-independent as a boundary condition.
The most important quantity in this expansion is $C_{AB}^{(1)}$, which encodes propagating modes in this theory. As a symmetric tensor on the sphere, it can be decomposed into a traceless tensor $C_{AB}$ and a trace part proportional to $\gamma_{AB}$.
The determinant condition (\ref{determinant}), however, implies at subleading order that the trace component is uniquely fixed by the scalar as
\begin{align}\label{decomposition}
	C_{AB}^{(1)}=C_{AB}-\gamma_{AB}\frac{ \varphi_1}{\varphi_0}\,,\qquad \gamma^{AB}C_{AB}=0.
\end{align}
The traceless part $C_{AB}$ encodes the spin-$2$ graviton mode as in GR, while the trace encodes the spin-$0$ scalar mode.

\subsection{Equations of motion at null infinity}\label{sec3}

In this section, we solve the field equations in the Jordan frame at future null infinity \(\mathscr{I}^{+}\) within the Bondi-Sachs gauge. In leading order, Equation~\eqref{EqScalar} implies
$\partial_u\varphi_0=0$, while the angular dependence of $\varphi_0$ is not constrained by the field equations. However, since $\varphi_0$ represents the effective gravitational constant in this theory, we impose as a boundary condition that $\varphi_0$ is a constant. In \(\mathcal{O}(r^{-2})\), we obtain
\begin{equation}\label{EqSubleading}
	2\partial_u \varphi_2 + D^A D_A \varphi_1 + \frac{f'_0}{2f_0}\partial_u\varphi_1^2 = 0.
\end{equation}
As typical in the Bondi-Sachs analysis, we reduce the equations to two-dimensional equations on the sphere, with the fixed metric $\gamma_{AB}$, which will be used to lower or raise indices and to define a covariant derivative $D_A\gamma_{BC}=0$.
The first two terms in \eqref{EqSubleading} are analogous to the asymptotic behavior of a minimally coupled Klein-Gordon field, while the last term is a non-linear term that originates from the non-minimal coupling in the Tachyon model.\\
Let us now turn to gravitational field equations and their consequences. The $rr$-component of  Eq.(\ref{EqEinstein}) implies that $\beta=\beta_1/r+\beta_2/r^2$ where
\begin{align}
	\beta_1 = -\frac{\varphi_1}{2\varphi_0},\qquad \beta_2 = -\frac{1}{32}C_{AB}C^{AB}-\frac{\varphi_1^2}{16\varphi_0^2}(\varphi_0f_0-1)-\frac{3}{4}\frac{\varphi_2}{\varphi_0}.
\end{align}
The fact that \(\beta_1\) is proportional to the leading scalar field perturbation is consistent with the fact that $\beta=O(r^{-2})$ in GR.

Next, the $ur$-component of (\ref{EqEinstein}), implies that
$ V = -r + 2m + \mathcal{O}(r^{-1})$, where \( m \) is the Bondi mass aspect and is a quasi-local measure of energy in GR. In the current scalar-tensor model, it appears that
\begin{equation}\label{jordanm}
	\mathcal{M}  =m-\frac{1}{8}\frac{f^{'}_0}{f_0}\partial_u\varphi_1^2,
\end{equation}
is a more natural quantity, which we call the modified mass aspect.
From the $uu$-component of (\ref{EqEinstein}), we obtain an evolution equation for the latter
\begin{equation}\label{bondimass}
	\partial_u\mathcal{M} = \frac{1}{4} D^A D^B \dot{{C}}_{AB} - \frac{1}{8} \dot{{C}}_{AB} \dot{{C}}^{AB} - (1 + \frac{f_0}{4})N^2,
\end{equation}
Hereafter, overdot denotes the partial derivative with respect to retarded time $u$, and $N \equiv \dot{\varphi}_1$ which may be called the scalar news. This equation governs the rate of change of the modified mass aspect $\mathcal{M}$, showing how the Tachyon field contributes to mass loss at null infinity. Defining the energy as the integral of $\mathcal{M}$ over the sphere, we observe that the linear term on the right-hand side vanishes, and we find that the modified mass aspect is strictly decreasing in the presence of the tensor or scalar news. This is a generalization of the Bondi mass loss formula to this model. Although the linear term right-hand and side of \eqref{bondimass} is irrelevant for energy, it plays a crucial role in the derivation of the memory effects as we will see below.

Next, the $rA$-component of (\ref{EqEinstein}) implies
\begin{equation}
	U^A = -\frac{1}{2 r^2} D_B C^{AB} + \frac{1}{2 r^2} \frac{D^A \varphi_1}{\varphi_0} - \frac{2\mathcal{N}^A}{3r^3} + \mathcal{O} \left( \frac{1}{r^4} \right),
\end{equation}
where $\mathcal{N}^A(u,x^A)$ in the subleading term is an integration constant, called the angular momentum aspect. It obeys an evolution equation implied by the $uA$-component of \eqref{EqEinstein}
\begin{align}\label{angmom}
\partial_u\mathcal{N}_A &= D_A \mathcal{M} +
 \frac{1}{4} \Big( D_B D_A D_C {C}^{BC} - D_B D^B D_C {C}^C_A + \frac{1}{2} D_A D^B D_B \varphi_1 \Big) \nonumber \\
& - \frac{1}{16} {D}_A \left( N_C^{\ B}{C}^C_{\ B} \right)
+ \frac{1}{4} N_C^{\ B} {D}_A{C}^C_{\ B}
+ \frac{1}{4} {D}_B \left( N^C_{\ A} {C}^B_{\ C} - {C}^C_{\ A} N_C^{\ B} \right) \nonumber \\
&+\frac{1+f_0}{8\varphi_0}(-3ND_A\varphi_1+\varphi
_1D_AN)+\frac{1}{4\varphi_0}\frac{f_0^{'}}{f_0}D_A\partial_u\varphi_1^2.
\end{align}
Similarly to \eqref{bondimass}, Eq.\eqref{angmom} contains linear terms in the shear $C_{AB}$ and the scalar, appearing in the first line, as well as quadratic terms.

\section{Gravitational memory effects}\label{sec4}
In the Tachyon model, the GWs consist of three distinct polarization modes collected in Eq.\eqref{decomposition}: the spin-2 transverse-traceless tensor modes $C_{AB}$, also present in GR, and a scalar breathing mode, driven by fluctuations in the Tachyon field $\varphi_1$. The time variation of these quantities, which we call the news, carries energy from the system, explicit in \eqref{bondimass}. GW's main feature is an oscillatory behavior. However, crucially, it can be shown that both $C_{AB},\varphi_1$ can contain a  memory mode defined as
\begin{align}
    \Delta C_{AB}(x^A)&\equiv \lim_{u\to\infty}\big( C_{AB}(u,x^A)-C_{AB}(-u,x^A)\big)\,,\\ \Delta \varphi_1(x^A)&\equiv \lim_{u\to\infty}\big( \varphi_1(u,x^A)-\varphi_1(-u,x^A)\big),
\end{align}
\ie, a nonzero shift in the waveform.
In this section, we address several aspects of memory effects in the Tachyon model. First, we show that the Einstein equations derived in section \ref{sec3} imply that the memory is generically nonzero and dictated by the evolution of the system. Next, we point out a non-trivial observable associated with the memory, the displacement effect.

\subsection{Memory from Bondi evolution equations}\label{subs4.1}
The Bondi shear $C_{AB}$ can be decomposed in terms of two scalars $C^\pm(u,x^A)$ with electric and magnetic parity \cite{Faye:2024utu} (angle brackets denote the symmetric trace-free part)
\begin{align}\label{parity decomp}
     C_{AB} &= D_{\langle A} D_{B\rangle} \, C^+ + \varepsilon^C_{\;A} D_{\langle B} D_{C \rangle} \,C^- \,.
\end{align}
The Bondi mass loss equation determines the memory in the electric sector. To see this, integrate \eqref{bondimass} over time and rearrange to find
\begin{align}\label{memory mass}
	\frac{1}{4} D^A D^B \Delta{C}_{AB}&=\Delta\mathcal{M} +\mathcal{F}\, ,
\end{align}
where the total flux of energy at a given point on the celestial sphere is given by
\begin{align}
    \mathcal{F}(x^A)&=\frac{1}{8} \int_{-\infty}^\infty du\left[\dot{{C}}_{AB} \dot{{C}}^{AB} - (1 + \frac{f_0}{4})N^2\right].
\end{align}
Using the decomposition \eqref{parity decomp} in \eqref{memory mass}, we find is an equation for $\Delta C^+$
\begin{align}\label{memory electric}
    \frac{1}{8}D^2(D^2+2)\Delta C^+=\Delta\mathcal{M} +\mathcal{F},
\end{align}
which can be inverted using a Green function integral \cite{Perez:2023uwt}.

Equation \eqref{memory electric} implies that in a dynamical situation, the electric memory $\Delta C^+$ is sourced by two elements: I) the change in the mass aspect $\Delta\mathcal{M}$, which characterizes the difference between the initial and final states of the system at future and past timelike infinity; II) the energy flux carried away by massless radiation at null infinity. The former is called the linear or ordinary memory in some references, while the latter is referred to as the nonlinear or null memory.

One may wonder if memory effects in the magnetic part of the shear $\Delta C^-$, as well as the breathing memory $\Delta \varphi_1$ can also be deduced from balance equations. The natural candidates are \eqref{angmom},\eqref{EqSubleading} respectively. However, we notice that their structure are different from that of \eqref{bondimass}. The linear term in the shear in \eqref{angmom} as well as the linear term in $\varphi_1$ in \eqref{EqSubleading} contain no time derivatives, unlike \eqref{bondimass}. Let us first focus on the simpler equation \eqref{EqSubleading}. Subtracting this equation at $u\to \pm\infty$ where $N,N_{AB}\to 0$, we obtain
\begin{align}
    D^2\Delta\varphi_1&=\Delta(\dot{\varphi_2})
\end{align}
This equation states that the breathing meomry $\Delta\varphi_1$ is nonvanishing iff $\varphi_2$ grows linearly as $u\to \infty$ or $u\to -\infty$. An asymptotic analysis at timelike infinity similar to \cite{Perez:2023uwt} reveals that this is the case. We leave a careful analysis of the Tachyon model at timelike/spatial infinity to a later work.

Now we turn to \eqref{angmom}, which can be decomposed into gradient and curl parts. The latter, which can be extracted by applying $\varepsilon^{AD}D_D$ on both sides of \eqref{angmom} reveals
\begin{align}
     \varepsilon^{AD}D_D\dot{\mathcal{N}}_A=\frac{1}{2} \varepsilon_{AD}D^DD_B D^{[A} D_C {C}^{B]C}
\end{align}
Expanding the right hand side using \eqref{parity decomp}, and subtracting this equation at $u\to \pm \infty$ reveals that the magnetic memory $\Delta C^-$ is nonvanishing if and only if (the magnetic component of) the angular momentum aspect grows linearly in $u$ at late or early times. However, such behavior is not present in the scattering of  particles~\cite{Saha:2019tub,Sahoo:2021ctw}, nor in the matching of null infinity with a consistent boundary condition at timelike infinity introduced in~\cite{Compere:2023qoa}.

\subsection{A memory observable: Displacement effect}\label{subs4.2}

We can strengthen this argument by considering the displacement induced by GWs on nearby test masses. This is most easily described by the geodesic deviation equation
\begin{equation}
	u^\gamma \nabla_\gamma (u^\beta \nabla_\beta X^\alpha) = -R^\alpha_{\;\beta \gamma \delta} u^\beta X^\gamma u^\delta,
\end{equation}
where $u^\alpha$ is the tangent vector to the reference geodesic followed by one of the test masses, $X^\alpha$ is the \textit{deviation} vector representing the separation between the test masses, such that $X\cdot u=0$,  and $R^\alpha_{\;\beta \gamma \delta}$ is the Riemann curvature tensor. To analyze this, we decompose $X^\alpha$ in an orthonormal tetrad $\{e^\alpha_{\hat{0}}, e^\alpha_{\hat{i}}\}$, with $e^\alpha_{\hat{0}} = u^\alpha$ aligned along the geodesic and $e^\alpha_{\hat{i}}$ (for $i = 1, 2, 3$) that spans the orthogonal spatial directions, hence $X^\alpha = X^{\hat{i}} e^\alpha_{\hat{i}}$. Substitution into the geodesic deviation equation yields
\begin{equation}
	\ddot{X}_{\hat{i}} = -R_{\hat{i}\hat{0} \hat{j} \hat{0}} X^{\hat{j}},
\end{equation}
with $\ddot{X}^{\hat{i}}$ as the second derivative with respect to proper time $\tau$ along the geodesic. For an observer with a large luminosity distance $r$, a convenient basis, expressed in Bondi-Sachs coordinates, is
\begin{align}
    e_{\hat{0}}=\partial_u+O(1/r)\,,\qquad e_{\hat{r}}=\partial_r-\partial_u+O(1/r)\,,\qquad e_{\hat{A}}=\frac{1}{r}E_{\hat{A}}^A\partial_A+O(1/r^2).
\end{align}
 What is nice about this basis is that $e_{\hat{0}}$ is asymptotic to a timelike geodesic, and the transverse directions are represented by $e_{\hat{A}}$. As a result, the geodesic deviation equation is split into one for $\ddot{X}_{\hat{r}}$ and one for $\ddot{X}_{\hat{A}}$. However, the former is subleading because the Riemann tensor contracted with $e_{\hat{r}}$ is subleading, while the leading effect appears in the transverse projection
\begin{equation}
	R_{\hat{0} \hat{A} \hat{0} \hat{B}} = -\frac{1}{2r} \left( \partial_u^2 {C}_{\hat{A}\hat{B}} - \delta_{\hat{A}\hat{B}} \frac{\partial_u^2\varphi_1}{\varphi_0} \right) + O(r^{-2}),
\end{equation}
where $C_{\hat{A}\hat{B}} = E_{\hat{A}}^A E_{\hat{B}}^B C_{AB}$. Finally, we note that asymptotically, $d/d\tau=d/du+O(1/r)$ and therefore the geodesic deviation can be integrated to
\begin{equation}\label{displacement}
	\Delta X_{\hat{A}} = \frac{1}{2r} \left( \Delta {C}_{\hat{A}\hat{B}} - \delta_{\hat{A}\hat{B}} \frac{ \Delta\varphi_1}{\varphi_{0}}\right) X^{\hat{B}}_0 + O(r^{-2}),
\end{equation}
where $X^{\hat{B}}_0$ is the initial separation, and $\Delta f \equiv f(u) - f(u_0)$. The first term in the parentheses encodes the two $+,\times$ polarizations, while the second term is the so-called breathing mode. $\Delta X_{\hat{A}}(u)$ is the displacement caused by GWs as a function of time. In particular, taking $u$ to be large, when the GW has passed, we observe that there is a permanent displacement effect corresponding to each mode with memory.

The tensor and scalar modes have different observable effects. First of all, as visible from \eqref{displacement}, tensor modes cause shearing effects in a congruences of timelike geodesics, while the breathing mode creates isotropic expansion/contraction of the congruence (its name suggests). Also, they have different angular dependence on the celestial sphere: The tensor mode's multipole expansion starts at $\ell=2$, as well known from the quadrupole formula, while the scalar wave can have monople and dipole as well. However, since the monopole is time independent at leading post-Newtonian order, the leading contribution of the scalar radiation is \textit{dipolar} \cite{Favata:2008ti}.

\section{Soft theorems}\label{sec5}

In this section, we explore the role of soft theorems in the framework of the Tachyon model with non-minimal coupling to gravity. Soft theorems describe the behavior of scattering amplitudes when an external particle becomes soft, meaning that its momentum approaches zero. These theorems have been extensively studied in gauge theory and gravity, where they reveal deep connections between the infrared (IR) properties of quantum field theories and asymptotic symmetries of spacetime \cite{Weinberg:1965nx}.
One of the most fundamental results in this field is the Weinberg soft theorem for gravitons \cite{Weinberg:1964ew}, which controls the effect of the addition of an external particle with momentum $q\to 0$ to the scattering amplitude
\begin{equation}
	\mathcal{M}_{n+1} (p_1, \dots, p_n, q) \approx \left( S^{(0)} + S^{(1)} + S^{(2)} \right) \mathcal{M}_n (p_1, \dots, p_n),
\end{equation}
where the terms \( S^{(0)}, S^{(1)}, \) and \( S^{(2)} \) correspond to the leading, subleading, and subsubleading soft factors. In gravity, these expressions encode the universal properties of long-wavelength perturbations and their connection with asymptotic symmetries, such as the BMS group. Our goal is to consider how a massless Tachyon field, non-minimally coupled to gravity, modifies the standard soft graviton theorem. To achieve this goal, we first compute the vertex factors of the Tachyon and graviton and then analyze their soft behavior to derive the corresponding soft factor.

\subsection{Soft graviton}\label{subs5.2}

To analyze the interaction structure, we expand the Tachyon field perturbatively around a background solution
\begin{equation}\label{phiexpansion}
	\varphi\approx \varphi_0 +  \delta\varphi,
\end{equation}
where \( \varphi_0 \) is the background value and \( \delta\varphi \) represents the first-order perturbation. We also Taylor expand \( f(\varphi) \) as
\begin{equation}
	f(\varphi) \approx f_0+ f_0^{'}\delta\varphi+\cdots\,.
\end{equation}
By perturbing the metric as $g_{\mu\nu}=\eta_{\mu\nu}+\kappa h_{\mu\nu}$, and expanding the action (\ref{intractionaction}) to leading order in $\kappa$, we find the leading interaction between \( h_{\mu\nu} \) and \( \delta\varphi \)
\begin{equation}
	S_{\text{int}} \supset -\kappa\int d^4x  \left( \frac{3}{2\varphi_0} -  \frac{1}{2} \frac{f_0}{\varphi_0} \right) h^{\mu\nu} \partial_\mu \delta\varphi \partial_\nu \delta\varphi .
\end{equation}
This term explicitly shows the interaction between the graviton and the perturbed Tachyon field. Next, we use this result to analyze its contribution to the soft factor. The emission of a soft graviton from an external Tachyon line contributes as
\begin{equation}
	S^{(0)} = \kappa \sum_{i=1}^{n} \left( \frac{3}{2\varphi_0} - \frac{1}{2} \frac{f_0}{\varphi_0} \right) \frac{p_i^{\mu} p_i^{\nu} \epsilon_{\mu\nu}}{p_i \cdot q}.
\end{equation}
We note that the non-minimal coupling function \(f(\varphi)\) directly modifies the standard soft theorem by adding a correction that is proportional to \( (\frac{3}{2\varphi_0} - \frac{1}{2} \frac{f_0}{\varphi_0})\). The subleading soft factor arises from the next-to-leading order term in the expansion of the scattering amplitude with respect to the soft graviton momentum \( q^\mu \). In standard gravity, \( S^{(1)} \) is related to the angular momentum of the external particles, modified by the graviton’s polarization. For our Tachyon model, we must account for the non-minimal coupling’s effect on this term.  The subleading soft factor involves the first derivative of the vertex with respect to the graviton momentum, contracted with the polarization tensor \( \epsilon_{\mu\nu} \). After performing the expansion and applying the on-shell conditions for the external Tachyon lines (with momenta \( p_i^\mu \)), we find
\begin{equation}
	S^{(1)} = \kappa \sum_{i=1}^n \left( \frac{3}{2 \varphi_0} - \frac{1}{2} \frac{f_0}{\varphi_0} \right) \frac{p_i^\mu \epsilon_{\mu\nu} q_\sigma J_i^{\sigma\nu}}{p_i \cdot q},
\end{equation}
where \( J_i^{\sigma\nu} = p_i^\sigma \frac{\partial}{\partial p_{i\nu}} - p_i^\nu \frac{\partial}{\partial p_{i\sigma}} \) is the angular momentum operator acting on the \( i \)-th external particle. We observe again that the (only) effect of the non-minimal coupling on the emission of a soft graviton is the modulation by the coupling function \( f(\varphi) \).
The more interesting problem is to consider the emission of a soft scalar particle. This is what we study next.

\subsection{Soft Scalar Theorem}
\label{sec:soft_scalar}
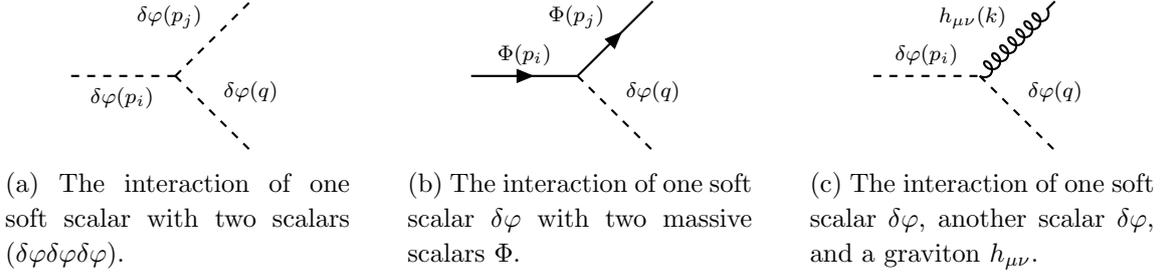
\begin{figure}[htbp]
  \centering

  \begin{subfigure}[b]{0.3\textwidth}
    \centering
   \begin{tikzpicture}
\begin{feynman}
\vertex (a);
\vertex [right=1.4cm of a] (b);
\vertex [above right=1.4cm of b] (c);
\vertex [below right=1.4cm of b] (d);
\diagram* {
(b) -- [scalar, thick,  color=black, edge label=\(\scriptstyle\delta \varphi(p_i)\)] (a),
(b) --  [scalar, thick, color=black, edge label=\(\scriptstyle\delta \varphi(p_j)\)](c),
(b) --  [scalar, thick,  color=black, edge label=\(\scriptstyle\delta \varphi(q)\)] (d),
};
\end{feynman}
\end{tikzpicture}
\caption{The interaction of one soft scalar with two scalars (\(\delta \varphi \delta \varphi \delta \varphi\)).}
  \end{subfigure}
  \hfill
  \begin{subfigure}[b]{0.3\textwidth}
    \centering
    \begin{tikzpicture}
\begin{feynman}
\vertex (a);
\vertex [right=1.4cm of a] (b);
\vertex [above right=1.4cm of b] (c);
\vertex [below right=1.4cm of b] (d);
\diagram* {
(a) -- [fermion, thick,  color=black, edge label=\(\scriptstyle\Phi(p_i)\)] (b),
(b) -- [fermion, thick,  color=black, edge label=\(\scriptstyle\Phi(p_j)\)] (c),
(b) -- [scalar, thick,  color=black, edge label=\(\scriptstyle\delta \varphi(q)\)] (d),
};
\end{feynman}
\end{tikzpicture}
\caption{The interaction of one soft scalar \(\delta \varphi\) with two massive scalars \(\Phi\).}
\label{fig:scalar_massive}
  \end{subfigure}
  \hfill
  \begin{subfigure}[b]{0.3\textwidth}
    \centering
    \begin{tikzpicture}
\begin{feynman}
\vertex (a);
\vertex [right=1.4cm of a] (b);
\vertex [above right=1.4cm of b] (c);
\vertex [below right=1.4cm of b] (d);
\diagram* {
(a) -- [scalar, thick,  color=black, edge label=\(\scriptstyle\delta \varphi(p_i)\)] (b),
(b) -- [gluon, thick,  color=black, edge label=\(\scriptstyle h_{\mu \nu}(k)\)] (c),
(b) -- [scalar, thick,  color=black, edge label=\(\scriptstyle\delta \varphi(q)\)] (d),
};
\end{feynman}
\end{tikzpicture}
\caption{The interaction of one soft scalar \(\delta \varphi\), another scalar \(\delta \varphi\), and a graviton \(h_{\mu \nu}\).}
\label{fig:scalar_graviton}
  \end{subfigure}
 \caption{Feynman diagrams contributing to soft theorems in the non-minimally coupled Tachyon theory.}
  \label{fig:soft-diagrams}
\end{figure}

In this subsection, we derive the soft theorem for the massless Tachyon perturbation \(\delta \varphi\), focusing on its scattering amplitudes in the soft limit (\(q \rightarrow 0\)). We consider interactions involving the soft scalar with itself, a massive scalar field \(\Phi\) (representing matter), and a graviton, all within the Einstein frame defined by the action (\ref{action}). The external soft massless scalar can appear in three vertices: (a) three Tachyon scalars, (b) one Tachyon scalar and two massive scalars, and (c) two massless scalars and a graviton. The Feynman diagrams for these cases are shown in Fig. \ref{fig:soft-diagrams}.

\subsubsection{Three Scalar Interaction}
The cubic interaction for the Tachyon perturbation \(\delta \varphi\) arises from non-linear terms in the action after expanding \(\varphi = \varphi_0 +  \delta \varphi\). In the Einstein frame, the interaction term is
\begin{equation}
    S_{\text{int}} \supset -\int d^4 x  \left( \frac{3}{2 \varphi_0^2} + \frac{1}{2} \left( \frac{f'_0}{\varphi_0} - \frac{f_0}{\varphi_0^2} \right) \right) \delta \varphi \partial_\mu \delta \varphi \partial^\mu \delta \varphi.
\end{equation}

The momentum-space vertex for three scalars with momenta \(p_i\), \(p_j\), and \(p_k\) (with \(p_k = q\) for the soft scalar) is
\begin{equation}
V_3 \sim  \left( \frac{3}{2 \varphi_0^2} + \frac{1}{2} \left( \frac{f'_0}{\varphi_0} - \frac{f_0}{\varphi_0^2} \right) \right) (p_i \cdot p_j).
\end{equation}
For the scattering amplitude \(\mathcal{M}_{n+1}(p_1, \ldots, p_n, q)\), attaching the soft scalar \(\delta \varphi(q)\) to the \(i\)-th external line (\(p_i\)), the propagator is
\begin{equation}
\frac{1}{(p_i + q)^2} \approx \frac{1}{2 p_i \cdot q},
\end{equation}
since \(p_i^2 = 0\) (massless Tachyon) and \(q^2 = 0\). The leading soft factor is
\begin{equation}
S_{\text{scalar-3$\varphi$}}^{(0)} = \left( \frac{3}{2 \varphi_0^2} + \frac{1}{2} \left( \frac{f'_0}{\varphi_0} - \frac{f_0}{\varphi_0^2} \right) \right) \sum_{i=1}^n \frac{p_i^2}{p_i \cdot q} = 0,
\end{equation}
due to \(p_i^2 = 0\).
Therefore, the contribution to the leading soft factor from this vertex is zero. Here, the summation over \( i \) runs over all external scalar particles involved in the hard process, i.e., those participating in the original \(n\)-point amplitude before the emission of the soft scalar.
\subsubsection{Soft Scalar with Massive Scalar Field}
To model interactions with matter, we introduce a massive scalar field \(\Phi\) with mass \(m_\Phi \neq 0\). The action in the Jordan frame is
\begin{equation}
S_m = \int d^4 x \sqrt{-g} \left[ -\frac{1}{2} g^{\mu \nu} \partial_\mu \Phi \partial_\nu \Phi - \frac{1}{2} m_\Phi^2 \Phi^2 \right].
\end{equation}
In the Einstein frame (\(\tilde{g}_{\mu \nu} = \varphi g_{\mu \nu}\)), this becomes
\begin{equation}
S_m = \int d^4 x \sqrt{-\tilde{g}} \left[ -\frac{1}{2} \varphi^{-1} \tilde{g}^{\mu \nu} \partial_\mu \Phi \partial_\nu \Phi - \frac{1}{2} m_\Phi^2 \varphi^{-2} \Phi^2 \right].
\end{equation}
Using (\ref{phiexpansion}), and
\begin{equation*}
\varphi^{-1} \approx \varphi_0^{-1} - \varphi_0^{-2} \delta \varphi, \quad \varphi^{-2} \approx \varphi_0^{-2} - 2 \varphi_0^{-3}  \delta \varphi,
\end{equation*}
yields the interaction Lagrangian
\begin{equation}
\mathcal{L}_{\text{int}} \supset \sqrt{-\tilde{g}} \left[ \frac{ \varphi_0^{-2}}{2} \delta \varphi \partial_\mu \Phi \partial^\mu \Phi + m_\Phi^2 \varphi_0^{-3}  \delta \varphi \Phi^2 \right].
\end{equation}
The cubic vertex for one soft scalar \(\delta \varphi(q)\) and two massive scalars \(\Phi(p_i, p_j)\) is
\begin{equation}
V_3 \sim  \left( \frac{\varphi_0^{-2}}{2} (p_i \cdot p_j) + m_\Phi^2 \varphi_0^{-3} \right).
\end{equation}
For the scattering amplitude, attaching \(\delta \varphi(q)\) to an external \(\Phi(p_i)\), we have
\begin{equation}
\mathcal{M}_{n+1} \supset  \sum_{i=1}^n \left[ \frac{\frac{\varphi_0^{-2}}{2} (p_i \cdot (p_i - q)) + m_\Phi^2 \varphi_0^{-3}}{(p_i - q)^2 - m_\Phi^2} \right] \mathcal{M}_n.
\end{equation}
Since \(p_i \cdot (p_i - q) = m_\Phi^2 - p_i \cdot q\) and \((p_i - q)^2 - m_\Phi^2 = -2 p_i \cdot q\), the amplitude becomes
\begin{equation}
\mathcal{M}_{n+1} \approx  \sum_{i=1}^n \frac{\frac{\varphi_0^{-2}}{2} (m_\Phi^2 - p_i \cdot q) + m_\Phi^2 \varphi_0^{-3}}{-2 p_i \cdot q} \mathcal{M}_n.
\end{equation}
The leading soft factor is
\begin{equation}
S_{\text{scalar-massive}}^{(0)} =  - \frac{ m_\Phi^2}{4 \varphi_0^2} (\varphi_0 + 2) \sum_{i=1}^n \frac{1}{p_i \cdot q}.
\end{equation}
This non-zero soft factor introduces infrared divergences due to the \(\frac{1}{p_i \cdot q}\) pole, requiring regularization, and contrasts with the massless case. Here, the summation over \( i \) runs over all external massive scalar particles \( \Phi \) present in the scattering process. Each such particle can emit the soft scalar through the direct cubic interaction term.

\subsubsection{Soft Scalar with Scalar and Graviton}
The interaction term for two scalars (\(\delta \varphi\)) and a graviton (\(h_{\mu \nu}\)) is
\begin{equation}
\mathcal{L}_{\text{int}} \supset -\kappa  \left( \frac{3}{2 \varphi_0} - \frac{1}{2} \frac{f_0}{\varphi_0} \right) h^{\mu \nu} \partial_\mu \delta \varphi \partial_\nu \delta \varphi.
\end{equation}
The vertex factor for two scalars with momenta \(p_i\) and \(q\) (soft scalar), and a graviton with momentum \(k\) and polarization \(\epsilon_{\mu \nu}\), is
\begin{equation}
V_3 = i \kappa \left( \frac{3}{2 \varphi_0} - \frac{1}{2} \frac{f_0}{\varphi_0} \right) (p_i^\mu q^\nu + p_i^\nu q^\mu) \epsilon_{\mu \nu}.
\end{equation}
The scattering amplitude for \(n\) scalars plus one soft scalar interacting with a graviton is
\begin{equation}
\mathcal{M}_{n+1} \supset \sum_{i=1}^n \left[ i \kappa \left( \frac{3}{2 \varphi_0} - \frac{1}{2} \frac{f_0}{\varphi_0} \right) (p_i^\mu q^\nu + p_i^\nu q^\mu) \epsilon_{\mu \nu} \cdot \frac{1}{2 p_i \cdot q} \right] \mathcal{M}_n.
\end{equation}
Simplifying, the soft factor is
\begin{equation}
S_{\text{scalar-graviton}}^{(0)} = -\frac{\kappa}{2} \left( \frac{3}{2 \varphi_0} - \frac{1}{2} \frac{f_0}{\varphi_0} \right) \sum_{i=1}^n \frac{(p_i^\mu q^\nu + p_i^\nu q^\mu) \epsilon_{\mu \nu}}{p_i \cdot q}.
\end{equation}
In this expression, the summation runs over all external massless scalar particles \( \delta\varphi \). Each such scalar contributes to the soft factor through the scalar-graviton vertex, where the soft scalar is emitted in the presence of a graviton.

\subsubsection{Total soft scalar factor}
Summing the contributions from the three interactions, the total scattering amplitude in the soft limit is
\begin{equation}
\mathcal{M}_{n+1} = \left[ S_{\text{scalar-3$\varphi$}}^{(0)} + S_{\text{scalar-massive}}^{(0)} + S_{\text{scalar-graviton}}^{(0)} \right] \mathcal{M}_n.
\end{equation}
Assuming \(\kappa   = 1\), the total soft factor is
\begin{equation}
S_{\text{scalar}}^{(0)} =
- \frac{m_\Phi^2}{4\varphi_0^2}(\varphi_0 + 2)
\sum_{i=1}^n \frac{1}{p_i \cdot q}
- \frac{1}{2} \left( \frac{3}{2\varphi_0} - \frac{1}{2} \frac{f_0}{\varphi_0} \right)
\sum_{i=1}^n \frac{(p_i^\mu q^\nu + p_i^\nu q^\mu) \epsilon_{\mu \nu}}{p_i \cdot q}.
\end{equation}
Note that the summations in the first and second terms of the soft factor run over different sets of external particles: the first term involves all external massive scalar fields \( \Phi \), from which the soft scalar is emitted via direct coupling, whereas the third term involves all external massless scalar fields \( \delta\varphi \), from which the soft scalar is emitted through a scalar-graviton interaction vertex.

\section{Conclusions}\label{sec6}
In this work, we studied gravitational memory effects and soft theorems in a massless Tachyon scalar field model that has a non-minimal coupling to gravity. Using the Bondi-Sachs framework, we consider the asymptotic behavior of the Tachyon field and metric near the null infinity and derive the corresponding field equations in this regime. We found that (the electric component of) the tensor mode and the scalar mode contribute to permanent distortions in the geometry of spacetime, usually known as gravitational memory effects.
The former is captured by the shear of the outgoing null geodesics also appears in GR. The scalar field produces a novel expansion effect in the congruence of outgoing null geodesics, an isotropic breathing mode. Moreover, the presence of the Tachyon scalar modifies the standard Bondi mass loss formula, indicating the emission of energy from the system to null infinity. The scalar radiation is even more leading in the multipolar expansion~\cite{Bernard:2022noq}.
We also considered the observable consequences of these memory effects through the geodesic deviation equation. We showed that there are distinct signatures for tensor and scalar modes in the permanent displacement of the test particles.

Given the correspondence between soft theorems and memory effects, we analyzed soft theorems for gravitons and scalar perturbations in this model. The results show that the non-minimum coupling function of the Tachyon field changes the leading and subleading soft graviton coefficients.
The soft scalar can appear in three vertices, shown in Fig.~\ref{fig:soft-diagrams}. We found that the contribution to the leading soft theorem vanishes in the vertex involving three Tachyon scalars, while nonzero contributions come from vertices involving massive fields or gravitons. The resulting soft scalar theorem shows that the diagram diverges as the inverse of the frequency in the soft limit. This is consistent with the results on the memory effect found in section \ref{sec4}.




\bibliographystyle{JHEP}
\bibliography{biblio}

\providecommand{\href}[2]{#2}\begingroup\raggedright\begin{thebibliography}{10}

\bibitem{Zeldovich:1974gvh}
Y.B.~Zel'dovich and A.G.~Polnarev, \emph{{Radiation of gravitational waves by a
  cluster of superdense stars}}, {\emph{Sov. Astron.} {\bfseries 18} (1974)
  17}.

\bibitem{Braginsky:1985vlg}
V.B.~Braginsky and L.P.~Grishchuk, \emph{{Kinematic Resonance and Memory Effect
  in Free Mass Gravitational Antennas}}, {\emph{Sov. Phys. JETP} {\bfseries 62}
  (1985) 427}.

\bibitem{braginsky1987gravitational}
V.B.~Braginsky and K.S.~Thorne, \emph{Gravitational-wave bursts with memory and
  experimental prospects}, {\emph{Nature} {\bfseries 327} (1987) 123}.

\bibitem{Christodoulou:1991cr}
D.~Christodoulou, \emph{{Nonlinear nature of gravitation and gravitational wave
  experiments}}, \href{https://doi.org/10.1103/PhysRevLett.67.1486}{\emph{Phys.
  Rev. Lett.} {\bfseries 67} (1991) 1486}.

\bibitem{Thorne:1992sdb}
K.S.~Thorne, \emph{{Gravitational-wave bursts with memory: The Christodoulou
  effect}}, \href{https://doi.org/10.1103/PhysRevD.45.520}{\emph{Phys. Rev. D}
  {\bfseries 45} (1992) 520}.

\bibitem{Zhang:2017geq}
P.M.~Zhang, C.~Duval, G.W.~Gibbons and P.A.~Horvathy, \emph{{Soft gravitons and
  the memory effect for plane gravitational waves}},
  \href{https://doi.org/10.1103/PhysRevD.96.064013}{\emph{Phys. Rev. D}
  {\bfseries 96} (2017) 064013}
  [\href{https://arxiv.org/abs/1705.01378}{{\ttfamily 1705.01378}}].

\bibitem{Zhang:2017rno}
P.M.~Zhang, C.~Duval, G.W.~Gibbons and P.A.~Horvathy, \emph{{The Memory Effect
  for Plane Gravitational Waves}},
  \href{https://doi.org/10.1016/j.physletb.2017.07.050}{\emph{Phys. Lett. B}
  {\bfseries 772} (2017) 743}
  [\href{https://arxiv.org/abs/1704.05997}{{\ttfamily 1704.05997}}].

\bibitem{Flanagan:2018yzh}
E.E.~Flanagan, A.M.~Grant, A.I.~Harte and D.A.~Nichols, \emph{{Persistent
  gravitational wave observables: general framework}},
  \href{https://doi.org/10.1103/PhysRevD.99.084044}{\emph{Phys. Rev. D}
  {\bfseries 99} (2019) 084044}
  [\href{https://arxiv.org/abs/1901.00021}{{\ttfamily 1901.00021}}].

\bibitem{Seraj:2021rxd}
A.~Seraj and B.~Oblak, \emph{{Gyroscopic Gravitational Memory}},
  \href{https://arxiv.org/abs/2112.04535}{{\ttfamily 2112.04535}}.

\bibitem{Seraj:2022qyt}
A.~Seraj and B.~Oblak, \emph{{The Precession Caused by Gravitational Waves}},
  \href{https://doi.org/10.1103/PhysRevLett.129.061101}{\emph{Phys. Rev. Lett.}
  {\bfseries 129} (2022) 061101}
  [\href{https://arxiv.org/abs/2203.16216}{{\ttfamily 2203.16216}}].

\bibitem{Pasterski:2015zua}
S.~Pasterski, \emph{{Asymptotic Symmetries and Electromagnetic Memory}},
  \href{https://doi.org/10.1007/JHEP09(2017)154}{\emph{JHEP} {\bfseries 09}
  (2017) 154} [\href{https://arxiv.org/abs/1505.00716}{{\ttfamily
  1505.00716}}].

\bibitem{Seraj:2022qqj}
A.~Seraj and T.~Neogi, \emph{{Memory effects from holonomies}},
  \href{https://arxiv.org/abs/2206.14110}{{\ttfamily 2206.14110}}.

\bibitem{Strominger:2017zoo}
A.~Strominger, \emph{{Lectures on the Infrared Structure of Gravity and Gauge
  Theory}},  \href{https://arxiv.org/abs/1703.05448}{{\ttfamily 1703.05448}}.

\bibitem{DeLuca:2024bpt}
V.~De~Luca, J.~Khoury and S.S.C.~Wong, \emph{{Gravitational memory and soft
  theorems: the local perspective}},
  \href{https://arxiv.org/abs/2412.01910}{{\ttfamily 2412.01910}}.

\bibitem{Barack:2018yly}
L.~Barack et~al., \emph{{Black holes, gravitational waves and fundamental
  physics: a roadmap}},
  \href{https://doi.org/10.1088/1361-6382/ab0587}{\emph{Class. Quant. Grav.}
  {\bfseries 36} (2019) 143001}
  [\href{https://arxiv.org/abs/1806.05195}{{\ttfamily 1806.05195}}].

\bibitem{Hou:2021oxe}
S.~Hou, T.~Zhu and Z.-H.~Zhu, \emph{{Asymptotic analysis of Chern-Simons
  modified gravity and its memory effects}},
  \href{https://doi.org/10.1103/PhysRevD.105.024025}{\emph{Phys. Rev. D}
  {\bfseries 105} (2022) 024025}
  [\href{https://arxiv.org/abs/2109.04238}{{\ttfamily 2109.04238}}].

\bibitem{Bernard:2022noq}
L.~Bernard, L.~Blanchet and D.~Trestini, \emph{{Gravitational waves in
  scalar-tensor theory to one-and-a-half post-Newtonian order}},
  \href{https://doi.org/10.1088/1475-7516/2022/08/008}{\emph{JCAP} {\bfseries
  08} (2022) 008} [\href{https://arxiv.org/abs/2201.10924}{{\ttfamily
  2201.10924}}].

\bibitem{Gorji:2022hyy}
M.A.~Gorji, T.~Matsuda and S.~Mukohyama, \emph{{Cosmological memory effect in
  scalar-tensor theories}},
  \href{https://doi.org/10.1103/PhysRevD.106.024013}{\emph{Phys. Rev. D}
  {\bfseries 106} (2022) 024013}
  [\href{https://arxiv.org/abs/2204.09182}{{\ttfamily 2204.09182}}].

\bibitem{Heisenberg:2023prj}
L.~Heisenberg, N.~Yunes and J.~Zosso, \emph{{Gravitational wave memory beyond
  general relativity}},
  \href{https://doi.org/10.1103/PhysRevD.108.024010}{\emph{Phys. Rev. D}
  {\bfseries 108} (2023) 024010}
  [\href{https://arxiv.org/abs/2303.02021}{{\ttfamily 2303.02021}}].

\bibitem{Trestini:2023khz}
D.~Trestini, \emph{{Gravitational radiation of compact binary systems in
  general relativity and in scalar-tensor theories}}, Ph.D. thesis, Institut
  d'Astrophysique de Paris, France, Sorbonne Universit{\'e}, 2023.

\bibitem{BenAchour:2024zzk}
J.~Ben~Achour, M.A.~Gorji and H.~Roussille, \emph{{Nonlinear gravitational
  waves in Horndeski gravity: scalar pulse and memories}},
  \href{https://doi.org/10.1088/1475-7516/2024/05/026}{\emph{JCAP} {\bfseries
  05} (2024) 026} [\href{https://arxiv.org/abs/2401.05099}{{\ttfamily
  2401.05099}}].

\bibitem{Hou:2020tnd}
S.~Hou and Z.-H.~Zhu, \emph{{Gravitational memory effects and
  Bondi-Metzner-Sachs symmetries in scalar-tensor theories}},
  \href{https://doi.org/10.1007/JHEP01(2021)083}{\emph{JHEP} {\bfseries 01}
  (2021) 083} [\href{https://arxiv.org/abs/2005.01310}{{\ttfamily
  2005.01310}}].

\bibitem{Tahura:2020vsa}
S.~Tahura, D.A.~Nichols, A.~Saffer, L.C.~Stein and K.~Yagi, \emph{{Brans-Dicke
  theory in Bondi-Sachs form: Asymptotically flat solutions, asymptotic
  symmetries and gravitational-wave memory effects}},
  \href{https://doi.org/10.1103/PhysRevD.103.104026}{\emph{Phys. Rev. D}
  {\bfseries 103} (2021) 104026}
  [\href{https://arxiv.org/abs/2007.13799}{{\ttfamily 2007.13799}}].

\bibitem{Seraj:2021qja}
A.~Seraj, \emph{{Gravitational breathing memory and dual symmetries}},
  \href{https://doi.org/10.1007/JHEP05(2021)283}{\emph{JHEP} {\bfseries 05}
  (2021) 283} [\href{https://arxiv.org/abs/2103.12185}{{\ttfamily
  2103.12185}}].

\bibitem{Tahura:2025ebb}
S.~Tahura, D.A.~Nichols and K.~Yagi, \emph{{Gravitational-wave memory effects
  in the Damour-Esposito-Far\`ese extension of Brans-Dicke theory}},
  \href{https://arxiv.org/abs/2501.07488}{{\ttfamily 2501.07488}}.

\bibitem{Koyama:2020vfc}
K.~Koyama, \emph{{Testing Brans-Dicke Gravity with Screening by Scalar
  Gravitational Wave Memory}},
  \href{https://doi.org/10.1103/PhysRevD.102.021502}{\emph{Phys. Rev. D}
  {\bfseries 102} (2020) 021502}
  [\href{https://arxiv.org/abs/2006.15914}{{\ttfamily 2006.15914}}].

\bibitem{Bose:2009zzd}
S.K.~Bose, \emph{{Aspects of tachyon theory}},
  \href{https://doi.org/10.1088/1742-6596/196/1/012022}{\emph{J. Phys. Conf.
  Ser.} {\bfseries 196} (2009) 012022}.

\bibitem{Calcagni:2006ge}
G.~Calcagni and A.R.~Liddle, \emph{{Tachyon dark energy models: dynamics and
  constraints}}, \href{https://doi.org/10.1103/PhysRevD.74.043528}{\emph{Phys.
  Rev. D} {\bfseries 74} (2006) 043528}
  [\href{https://arxiv.org/abs/astro-ph/0606003}{{\ttfamily
  astro-ph/0606003}}].

\bibitem{Sen:2002nu}
A.~Sen, \emph{{Rolling tachyon}},
  \href{https://doi.org/10.1088/1126-6708/2002/04/048}{\emph{JHEP} {\bfseries
  04} (2002) 048} [\href{https://arxiv.org/abs/hep-th/0203211}{{\ttfamily
  hep-th/0203211}}].

\bibitem{Fairbairn:2002yp}
M.~Fairbairn and M.H.G.~Tytgat, \emph{{Inflation from a tachyon fluid?}},
  \href{https://doi.org/10.1016/S0370-2693(02)02638-2}{\emph{Phys. Lett. B}
  {\bfseries 546} (2002) 1}
  [\href{https://arxiv.org/abs/hep-th/0204070}{{\ttfamily hep-th/0204070}}].

\bibitem{Sen:1998sm}
A.~Sen, \emph{{Tachyon condensation on the brane anti-brane system}},
  \href{https://doi.org/10.1088/1126-6708/1998/08/012}{\emph{JHEP} {\bfseries
  08} (1998) 012} [\href{https://arxiv.org/abs/hep-th/9805170}{{\ttfamily
  hep-th/9805170}}].

\bibitem{Arutyunov:2000pe}
G.~Arutyunov, S.~Frolov, S.~Theisen and A.A.~Tseytlin, \emph{{Tachyon
  condensation and universality of DBI action}},
  \href{https://doi.org/10.1088/1126-6708/2001/02/002}{\emph{JHEP} {\bfseries
  02} (2001) 002} [\href{https://arxiv.org/abs/hep-th/0012080}{{\ttfamily
  hep-th/0012080}}].

\bibitem{RezaeiAkbarieh:2018ijw}
A.~Rezaei~Akbarieh and Y.~Izadi, \emph{{Tachyon Inflation in Teleparallel
  Gravity}}, \href{https://doi.org/10.1140/epjc/s10052-019-6819-z}{\emph{Eur.
  Phys. J. C} {\bfseries 79} (2019) 366}
  [\href{https://arxiv.org/abs/1812.06649}{{\ttfamily 1812.06649}}].

\bibitem{Nozari:2013mba}
K.~Nozari and N.~Rashidi, \emph{{Some Aspects of Tachyon Field Cosmology}},
  \href{https://doi.org/10.1103/PhysRevD.88.023519}{\emph{Phys. Rev. D}
  {\bfseries 88} (2013) 023519}
  [\href{https://arxiv.org/abs/1306.5853}{{\ttfamily 1306.5853}}].

\bibitem{Piao:2002nh}
Y.-S.~Piao, Q.-G.~Huang, X.-m.~Zhang and Y.-Z.~Zhang, \emph{{Nonminimally
  coupled tachyon and inflation}},
  \href{https://doi.org/10.1016/j.physletb.2003.07.047}{\emph{Phys. Lett. B}
  {\bfseries 570} (2003) 1}
  [\href{https://arxiv.org/abs/hep-ph/0212219}{{\ttfamily hep-ph/0212219}}].

\bibitem{Madler:2016xju}
T.~M\"adler and J.~Winicour, \emph{{Bondi-Sachs Formalism}},
  \href{https://doi.org/10.4249/scholarpedia.33528}{\emph{Scholarpedia}
  {\bfseries 11} (2016) 33528}
  [\href{https://arxiv.org/abs/1609.01731}{{\ttfamily 1609.01731}}].

\bibitem{Faye:2024utu}
G.~Faye and A.~Seraj, \emph{{Gyroscopic gravitational memory from
  quasi-circular binary systems}},
  \href{https://doi.org/10.1088/1361-6382/ada339}{\emph{Class. Quant. Grav.}
  {\bfseries 42} (2025) 035005}
  [\href{https://arxiv.org/abs/2409.02624}{{\ttfamily 2409.02624}}].

\bibitem{Perez:2023uwt}
A.~P\'erez, S.~Prohazka and A.~Seraj, \emph{{Fracton Infrared Triangle}},
  \href{https://doi.org/10.1103/PhysRevLett.133.021603}{\emph{Phys. Rev. Lett.}
  {\bfseries 133} (2024) 021603}
  [\href{https://arxiv.org/abs/2310.16683}{{\ttfamily 2310.16683}}].

\bibitem{Saha:2019tub}
A.P.~Saha, B.~Sahoo and A.~Sen, \emph{{Proof of the classical soft graviton
  theorem in $D$ = 4}},
  \href{https://doi.org/10.1007/JHEP06(2020)153}{\emph{JHEP} {\bfseries 06}
  (2020) 153} [\href{https://arxiv.org/abs/1912.06413}{{\ttfamily
  1912.06413}}].

\bibitem{Sahoo:2021ctw}
B.~Sahoo and A.~Sen, \emph{{Classical soft graviton theorem rewritten}},
  \href{https://doi.org/10.1007/JHEP01(2022)077}{\emph{JHEP} {\bfseries 01}
  (2022) 077} [\href{https://arxiv.org/abs/2105.08739}{{\ttfamily
  2105.08739}}].

\bibitem{Compere:2023qoa}
G.~Comp\`ere, S.E.~Gralla and H.~Wei, \emph{{An asymptotic framework for
  gravitational scattering}},
  \href{https://doi.org/10.1088/1361-6382/acf5c1}{\emph{Class. Quant. Grav.}
  {\bfseries 40} (2023) 205018}
  [\href{https://arxiv.org/abs/2303.17124}{{\ttfamily 2303.17124}}].

\bibitem{Favata:2008ti}
M.~Favata, \emph{{Gravitational-wave memory revisited: memory from the merger
  and recoil of binary black holes}},
  \href{https://doi.org/10.1088/1742-6596/154/1/012043}{\emph{J. Phys. Conf.
  Ser.} {\bfseries 154} (2009) 012043}
  [\href{https://arxiv.org/abs/0811.3451}{{\ttfamily 0811.3451}}].

\bibitem{Weinberg:1965nx}
S.~Weinberg, \emph{{Infrared photons and gravitons}},
  \href{https://doi.org/10.1103/PhysRev.140.B516}{\emph{Phys. Rev.} {\bfseries
  140} (1965) B516}.

\bibitem{Weinberg:1964ew}
S.~Weinberg, \emph{{Photons and Gravitons in $S$-Matrix Theory: Derivation of
  Charge Conservation and Equality of Gravitational and Inertial Mass}},
  \href{https://doi.org/10.1103/PhysRev.135.B1049}{\emph{Phys. Rev.} {\bfseries
  135} (1964) B1049}.

\end{thebibliography}\endgroup



\end{document}